\documentclass[aps,pra,showpacs,twocolumn,amsmath,amssymb,superscriptaddress,footinbib]{revtex4} 

\usepackage[english]{babel}
\usepackage{latexsym}
\usepackage{graphics}
\usepackage{subfigure}
\usepackage{epsfig}

\newcommand{\ie}{{\it i.e.}~}

\newcommand{\ket}[1]{| #1 \rangle}
\newcommand{\bra}[1]{\langle #1 |}
\newcommand{\dyad}[2]{\ket{#1}\!\bra{#2}}

\newcommand{\Id}{\mathbf{1}}
\newcommand{\tr}{\mathrm{tr}}

\begin{document}

\title{Infinite qubit rings with maximal nearest neighbor entanglement: 
the Bethe ansatz solution}

\author{U.~V.~Poulsen} 
\affiliation{Department of Physics and
  Astronomy, University of Aarhus, DK-8000 \AA rhus C., Denmark}
\author{T.~Meyer} 
\affiliation{Institut f\"ur Theoretische Physik III,
  Heinrich-Heine-Universit\"at D\"usseldorf, D-40225 D\"usseldorf,
  Germany} 
\author{D.~Bru\ss}
\affiliation{Institut f\"ur Theoretische
  Physik III, Heinrich-Heine-Universit\"at D\"usseldorf, D-40225
  D\"usseldorf, Germany} 
\author{M.~Lewenstein}
\affiliation{ICREA and ICFO-Institut de Ci\`encies Fot\`oniques, Parc
  Mediterrani de la Tecnologia, E-08860 Castelldefels (Barcelona),
  Spain} 
\affiliation{Institut f\"ur Theoretische Physik,
  Universit\"at Hannover, D-30167 Hannover, Germany}
 
\date{\today}

\begin{abstract}
  We search for translationally invariant states of qubits on a ring
  that maximize the nearest neighbor entanglement. This problem was
  initially studied by O'Connor and Wootters [Phys. Rev. A {\bf 63},
  052302 (2001)]. We first map the problem to the search for the
  ground state of a spin 1/2 Heisenberg XXZ model. Using the exact
  Bethe ansatz solution in the limit $N\rightarrow\infty$, we prove
  the correctness of the assumption of O'Connor and Wootters that the
  state of maximal entanglement does not have any pair of neighboring
  spins ``down'' (or, alternatively spins ``up''). For sufficiently
  small fixed magnetization, however, the assumption does not hold: we
  identify the region of magnetizations for which the states that
  maximize the nearest neighbor entanglement necessarily contain pairs
  of neighboring spins ``down''.
\end{abstract}


\maketitle
\section{Introduction}

The investigation of the role of entanglement in quantum and classical phase
transitions, and more generally the role of entanglement in many-body
quantum systems is one of the hottest interdisciplinary areas on the
borders between quantum information, quantum optics, atomic,
molecular, and condensed matter physics. Initially the studies of
entanglement in many-body systems have been motivated by the
possibility of employing entanglement for quantum computation in
optical lattices \cite{optical_lattice}, or precision measurements
with Bose-Einstein condensates \cite{BEC}. Recently, several lines of 
research have been followed:
\begin{itemize}
\item Studies of local entanglement in spin systems
  \cite{Osterloh,Nielsen,salikh-qwerty,salikh-qwerty1,Indranidi}, and
  more generally in various many-body systems (such as linear chains, see
  for instance \cite{linchain,Bogol,wolflin}), with particular
  attention to the role of entanglement in phase transitions.
\item Studies of the entropy of blocks of spins and the related ``area
  law'' \cite{area1}, indicating weak entanglement of blocks, and
  effects of criticality
  \cite{salikh-qwerty,area2,wolflin,meti-pukur}.
\item Studies of localisable entanglement and entanglement correlation
  length that diverges at the critical point
  \cite{ent_length,Ciracnew} and majorizes the standard correlation
  length (see also \cite{kachpoka}). In particular, it has been shown
  that the localisable entanglement
  (cf.~\cite{hmmmm,amrao-achhi-re-bhai,jin04:_localiz_entang_spin_chain})
  is bounded from above by the entanglement of assistance
  \cite{Bennett-assistance} and from below by correlation functions.
  It follows directly from these bounds that one can define an
  entanglement correlation length that diverges in quantum critical
  systems.
\item Studies of multipartite entanglement in many-body
  systems~\cite{Dagmar}.
\item Studies of dynamics and generation of entanglement in many-body
  systems \cite{Briegel_orey_orey,Sougato,talk_dilo_Osterloh,amader,
    meti-pukur}. In particular, implementations of the ``one-way
  quantum computer'' and short range teleportation \cite{BBCJPW} of an
  unknown state has been proposed by using the dynamics of spin
  systems in
  Refs.~\cite{Briegel_orey_orey,Briegel_ebong,meti-pukur,Sougato,Sougato_ebong,dpt}.
\item Studies of quantum information and entanglement theory inspired
  numerical codes to simulate quantum systems \cite{numerical}.
\end{itemize}

A different approach to the study of entanglement in many-body systems
has been proposed in two papers by Wootters and coworkers
\cite{Wootters}. In these papers, instead of looking at a specific
Hamiltonian, the authors asked the fundamental question ``what is the
maximal entanglement between two neighboring sites of an entangled
ring with translational invariance?'' Here, an entangled ring is a
chain of spins with periodic boundary conditions. Due to the so-called 
``monogamy of entanglement'' it is impossible for a site to be maximally
entangled with both its neighbors: shared entanglement is always less
than maximal \cite{brusspra,oldwootters}.  In Ref.~\cite{Wootters} the
question of the upper limit for the nearest neighbor (NN) entanglement was
simplified by introducing two additional restrictions on the allowed
states:
\begin{itemize}
\item[(i)] The state of $N$ spins 1/2 is an eigenstate of the
	$z$-component of the total spin (i.e. it has a fixed number of
  ``down'' spins $p\le N/2$)~\footnote{Note that one can also allow the
    state to be an incoherent mixture of several states, each with a
    fixed $p$: Since we will be optimizing a convex function, such a
    mixture cannot be optimal. Thus restriction (i) can be replaced
    with the formally weaker demand that the density operator commutes
    with the $z$-component of the total spin~\cite{meyer}.}.
\item[(ii)] Neighboring spins cannot both be ``down''.
\end{itemize}
Obviously, one can equally well study the same problem in terms of
spins ``up'', when $N\ge p\ge N/2$.  Both restrictions are based
on an educated guess for the optimal states for the general problem.
O'Connor and Wootters (OW) solved the restricted optimization problem by
relating it to an effective Hamiltonian for the one-dimensional
ferromagnetic XY model, and found the maximal nearest-neighbor
concurrence (cf. Sec.~\ref{sec:var_con_form}) for given $N$ and $p$ to be
\begin{equation}
  C^{\text{max}}_\text{OW}(N,p)
  =
  \frac{
    2\sin\left(\frac{p\pi}{N-p}\right)
  }{
    N\sin\left(\frac\pi{N-p}\right)
  }
\label{eq:cwootters}\; .
\end{equation} 
For given $N$ and $p$, Eq. \eqref{eq:cwootters} provides a lower
bound for the problem without restriction (ii). It may, or may not
happen that $C$ can be increased by also allowing states where two
neighboring spins are ``down''. We have recently studied finite size
rings and found that for a fixed $p$ restriction (ii) tends to play a
less important role as $N$ is increased~\cite{meyer}: For $p$ close to
$N/2$ one can increase the concurrence significantly by dropping
restriction (ii), but for $p \alt N/3$ OW's result is the
optimal one. In fact, already in Ref.~\cite{Wootters} it was shown
that for all even $N$ the ground state of a Heisenberg spin 1/2
antiferromagnetic ring maximizes the concurrence among the zero
magnetization ($p=N/2$) states although it violates restriction (ii).

By optimizing \eqref{eq:cwootters} with respect to $p$ one obtains a
lower bound on the overall optimal concurrence, \ie without any
restrictions besides the translational invariance. In the limit
$N\rightarrow\infty$, the optimal number of spins ``down'' in
Eq.~\eqref{eq:cwootters} approaches $p_\text{opt}\approx 0.301\;N$.
This leads to an asymptotic value of $C_\text{OW}^\text{max}\approx 0.434$.
Although Ref.~\cite{Wootters} as well as our previous
work~\cite{meyer} showed evidence for optimality of this number,
whether it can be improved was, so far, an open problem.

Wolf, Verstraete, and Cirac have in Ref.~\cite{wolf03:_entan_frust}
directly related OW's type of problems of looking for translationally
invariant states that maximize local entanglement to the study of the
ground state of a suitably defined ``parent'' Hamiltonian. In this
paper we use this method and employ the known exact solution of the
corresponding parent Hamiltonian to prove rigorously that:
\begin{itemize}
\item[(A)]In the limit $N\to \infty$ the translationally invariant state that
  maximizes the NN entanglement without any restriction coincides with
  the state found by OW at the optimal value
  of $p\simeq 0.301 N$. This means that it is not a superposition of 
	states with different $p$ values and does not contain simultaneously
  neighboring spin ``up'' and neighboring spin ``down'' pairs. 
\item[(B)]For fixed $p$ sufficiently close to $N/2$, \ie for 
  sufficiently small magnetizations,  assumption (ii) is not
  correct: the states that maximize the nearest neighbor entanglement
  necessarily contain simultaneously pairs of neighboring ``up'' and
  ``down'' spins. In the limit $N\rightarrow\infty$ we identify
  rigorously an interval of $p/N$ for which this is the case and show
  strong numerical evidence that this interval is optimal.
\end{itemize}

Our paper is organized as follows. In Section~\ref{sec:var_con_form}
we apply the method of Ref.~\cite{wolf03:_entan_frust} and derive the
corresponding parent Hamiltonian for a $N$ qubit ring.  In
Section~\ref{sec:parent-hamiltonian} we show the connection with the
``classical'' papers of Yang and Yang on the XXZ model. In
Section~\ref{sec:phasediagram} we discuss briefly the regimes of
parameters of interest and show that the present problem concerns the
``difficult'' parameter region of the phase diagram. In
Section~\ref{sec:int_eq} we present the analysis based on the limit
$N\to \infty$ of the Bethe ansatz solutions. We derive here the basic
integral equation, the solution of which allows to calculate the
desired energy of the system in question. In Section~\ref{sec:solving}
the numerical results are discussed.  In Section~\ref{sec:pert_calc}
we rigorously prove that the states that were conjectured in
Ref.~\cite{Wootters} to maximize the NN entanglement and confirmed by
us, indeed provide the maximum of the NN entanglement for sufficiently
small values of $p$.  We identify the region of $p/N$ where the latter
statement does not hold.  We conclude in
Section~\ref{sec:conclusions}.  The short appendix contains simple
analytic bounds on optimal magnetic field for which the NN concurence
is maximal.

\section{Variational concurrence formula}
\label{sec:var_con_form}

In this paper we will use the concurrence as our entanglement measure.
The concurrence~\cite{defconc} is defined as
$C(\rho)=\max\{\lambda_1-\lambda_2-\lambda_3-\lambda_4,0\}$, where
$\lambda_1\ge\lambda_2\ge\lambda_3\ge\lambda_4\ge0$ are the square
roots of the eigenvalues of $\rho\tilde\rho$, and
$\tilde\rho=(\sigma_y\otimes\sigma_y)\rho^* (\sigma_y\otimes\sigma_y)$
is the spin-flipped density matrix. The optimization problems that we
consider are complicated by the nonlinearity of the concurrence as
function of the density matrix. In our previous work~\cite{meyer}, we
showed how the optimization problem with fixed $p$ can be
reformulated as finding the ground state energy for each in a family
of spin-chain Hamiltonians. This family is parameterized by a single
real parameter and the optimal concurrence is minus the lowest ground
state energy that occurs when this parameter is varied. In this way a
complicated non-linear problem in a high-dimensional space is replaced
by a series of linear problems and one final one-parameter
optimization.
 
The derivation in Ref.~\cite{meyer} did not cover the case where
superpositions of states with different $p$ are allowed. To treat that
case, we turn to Ref.~\cite{wolf03:_entan_frust} where the following
general formula for the concurrence for systems of two spins $1/2$ has
been derived:
\begin{equation}
  \label{eq:var_form_2qubit}
  C(\rho) 
  = 
  \max \left\{ 
    0
    ,
    - \inf_{\det X=1} 
    \tr \left[ 
      \rho \left( X \otimes X^\dagger \mathbf{F} \right)
    \right]
  \right\}
  .
\end{equation}
Here $X$ is an arbitrary $2\times 2$ matrix of determinant 1, while
$\mathbf{F}$ is the flip (or swap) operator, interchanging the two
qubits:
\begin{equation}
  \label{eq:def_F}
  \mathbf{F} 
  =
  \dyad{\!\!\uparrow\uparrow}{\uparrow\uparrow\!\!}
  +\dyad{\!\!\downarrow\downarrow}{\downarrow\downarrow\!\!}
  +\dyad{\!\!\downarrow\uparrow}{\uparrow\downarrow\!\!}
  +\dyad{\!\!\uparrow\downarrow}{\downarrow\uparrow\!\!}
  .
\end{equation}
A useful parameterization of $X$ is obtained using the singular
value decomposition~\cite{horn85:_matrix_analysis}:
\begin{equation}
  \label{eq:X_sing_val}
  X 
  =
  U
  \begin{bmatrix}
    it & 0 \\
    0 &\frac{1}{it}
  \end{bmatrix}
  V^\dagger
  ,
\end{equation}
where $t\in[-\infty,\infty]$, $U,V \in U(2)$ and $\det U \cdot \det
V^\dagger = 1$. In fact, from Eq.~(\ref{eq:var_form_2qubit}) it is
clear that we can restrict to $U,V \in SU(2)$. We now rewrite
\begin{multline}
  \label{eq:xx_rewrite}
  X \otimes X^\dagger \;\mathbf{F}
  \\
  =
  \left( U \otimes V \right)
  \left(
    \begin{bmatrix}
      it & 0 \\
      0 & \frac{1}{it}
    \end{bmatrix}
    \otimes
    \begin{bmatrix}
      -it & 0 \\
      0 & \frac{1}{-it}
    \end{bmatrix}
  \right)
  \left( V^\dagger \otimes U^\dagger \right)
  \mathbf{F}
  \\
  =
  \left( U \otimes V \right)
    \begin{bmatrix}
      t^2 & 0 & 0 & 0 \\
      0   & -1 & 0 & 0\\
      0   & 0 & -1 & 0\\
      0   & 0 & 0 & t^{-2}
    \end{bmatrix}
    \mathbf{F}
  \left( U \otimes V \right)^\dagger
  \\
  =
  \left( U \otimes V \right)
    \begin{bmatrix}
      t^2 & 0 & 0 & 0 \\
      0   & 0 & -1 & 0\\
      0   & -1 & 0 & 0\\
      0   & 0 & 0 & t^{-2}
    \end{bmatrix}
  \left( U \otimes V \right)^\dagger.
\end{multline}
Let us define the matrix in square brackets as $h(t^2)$, i.e.,
\begin{equation}
  \label{eq:def_h_braket}
  h(s)
  =
  s\dyad{\!\!\uparrow\uparrow}{\uparrow\uparrow\!\!}
  +\frac{1}{s}\dyad{\!\!\downarrow\downarrow}{\downarrow\downarrow\!\!}
  -\dyad{\!\!\downarrow\uparrow}{\uparrow\downarrow\!\!}
  -\dyad{\!\!\uparrow\downarrow}{\downarrow\uparrow\!\!}
  .
\end{equation}
where $s=t^2$. Then we can rewrite Eq.~(\ref{eq:var_form_2qubit}) as
\begin{multline}
  \label{eq:C_of_h}
  C(\rho)
  =
  \\
  \max\left\{
    0
    ,
    -\inf_{s,U,V} \tr\left[
      \left(U\otimes V\right)^\dagger
      \rho
      \left(U\otimes V\right) 
      h(s)
    \right]
  \right\}
  .
\end{multline}

Our goal is to maximize the concurrence over all $\rho$ that can occur
as nearest neighbor density matrices on a translationally invariant
ring. If we always had $U=V$, it is easy to see that we could drop the
infimum over $U \in SU(2)$ in Eq.~(\ref{eq:C_of_h}) since if
$\rho=\tr_{3\ldots N} \Gamma$ with $\Gamma$ translationally invariant then $(U\otimes U)^\dagger \rho (U\otimes
U)=\tr_{3\ldots N} (U\otimes\ldots\otimes
U)^\dagger\Gamma(U\otimes\ldots\otimes U)$ where $(U\otimes\ldots\otimes
U)^\dagger\Gamma(U\otimes\ldots\otimes U)$ is translationally
invariant as well. To do the same for $U\neq V$, we can use the fact
that $h(s)$ is symmetric in the two qubits:
\begin{equation}
  \tr\left[
    \left( U \otimes V \right)^\dagger \rho \left( U \otimes V \right) h(s)
  \right]
  =
  \tr\left[
    \tilde\rho h(s)
  \right]
\end{equation}
where
\begin{multline}
  \tilde\rho=
  \\
  \frac{1}{2}\left\{
    \left( U \otimes V \right)^\dagger \rho \left( U \otimes V \right) 
    +
    \left( V \otimes U \right)^\dagger \rho \left( V \otimes U \right) 
  \right\}
  .
\end{multline}
If $N$ is even and $\rho=\tr_{3\ldots N} \Gamma$ then $\tilde\rho$ is
a nearest neighbor density matrix belonging to the following
translationally invariant state:
\begin{equation}
  \label{eq:mix_gamma}
  \begin{split}
    \frac{1}{2}\bigg\{ &
    \phantom{+}\left([U \otimes V]
      \otimes \ldots \otimes 
      [U \otimes V]\right)^\dagger \times
    \\
    &\phantom{aabbccdd}\times \Gamma\times
    \left([U \otimes V] \otimes 
      \ldots \otimes [U \otimes V]\right)
    \\
    &+
    \left([V \otimes U] \otimes \ldots \otimes
      [V \otimes U]\right)^\dagger \times
    \\
    &\phantom{aaccbbdd}\times\Gamma\times
    \left([V \otimes U]\otimes \ldots \otimes 
      [V \otimes U]\right) \bigg\}
    .
  \end{split}
\end{equation}
If $N$ is odd, the above construction does not work: we cannot fit an
integer number of $U\otimes V$ terms on the ring. By placing as many
terms $U\otimes V$ as possible, and taking the translationally variant 
mixture of the
resulting state, $\tilde\rho$ can be approximated up to a factor
$1/N$. In the limit of $N\rightarrow\infty$ we can ignore this
correction.

\section{The parent Hamiltonian}
\label{sec:parent-hamiltonian}

In this section we will follow the approach of
Ref.~\cite{wolf03:_entan_frust} to derive the parent spin $1/2$ XXZ
Hamiltonian, \ie the Hamiltonian whose ground state maximizes the NN
concurrence. We also make the connection to the classical papers on
the XXZ model by Yang and Yang
\cite{yang66:_one_dimen_I,yang66:_one_dimen_II,yang66:_one_dimen_III}.

In the previous section we showed that in the limit $N\rightarrow\infty$
\begin{equation}
  \label{eq:c_as_inf}
  C^\text{max}(N)
  =
  \max_\rho C(\rho) 
  = 
  - \inf_{s,\rho} \tr[ \rho h(s) ]
\end{equation}
where $\rho=\tr_{3\ldots N} \Gamma$ for some translationally invariant
$\Gamma$ of $N$ spins. The two-spin Hamiltonian (\ref{eq:def_h_braket}) 
can be written in terms of Pauli matrices as
\begin{equation}
  \label{eq:def_h}
  \begin{split}
    h(s) 
    =& 
    \frac{s}{4}\left(\Id+\sigma_z\right)\left(\Id+\sigma_z\right)
    +\frac{1}{4s}\left(\Id-\sigma_z\right)\left(\Id-\sigma_z\right)
    \\
    &-
    \sigma_+\sigma_+ - \sigma_-\sigma_-
    \\
    =& 
    \frac{1}{2}
    \left\{
      -\sigma_x \sigma_x - \sigma_y \sigma_y
      +\frac{1}{2}\left(s+\frac{1}{s}\right)
      \left(\sigma_z\sigma_z+1\right)\right.
    \\
    &
    \left. +\frac{1}{2}\left(s-\frac{1}{s}\right)
      \left(\sigma_z\Id+\Id\sigma_z\right)
    \right\}
    ,
  \end{split}
\end{equation}
where $\sigma_\pm=\sigma_x\pm i\sigma_y$.
Instead of working with $\rho$ and $h(s)$ we can use the translational
invariance and work with $\Gamma$ and a Hamiltonian for the whole ring
obtained by taking (\ref{eq:def_h}) for each NN pair:
\begin{equation}
  \label{eq:def_H}
  \begin{split}
    H_\text{Wolf}(s)
    &=
    \\
    \frac{1}{2N}&\sum_{i=1}^{N} 
    \big\{
      -\sigma_x^{i}  \sigma_x^{i+1} - \sigma_y^{i}\sigma_y^{i+1}
      - \Delta(s) \sigma_z^{i}\sigma_z^{i+1}
    \\
    &\phantom{\sum_{i=1}^{N} \big\{ -\sigma_x^{i}  \sigma_x^{i+1}}
    -2\mathcal{H}(s)\sigma_z^{i}
      -\Delta(s)
    \big\},
  \end{split}
\end{equation}
where
\begin{gather}
  \label{eq:def_Delta_and_mathcalH}
  \Delta(s)=-\frac{1}{2}\left(s+\frac{1}{s}\right) 
  \qquad
  \mathcal{H}(s)=-\frac{1}{2}\left(s-\frac{1}{s}\right)
  .
\end{gather}
We have then reformulated the overall optimization problem as
\begin{equation}
  \label{eq:def_global_opt}
  C^\text{max}(N)
  =
  \max_\rho C(\rho)=-\inf_{s,\Gamma} \tr[ \Gamma H_\mathrm{Wolf}(s) ],
\end{equation}
where $\rho$ is restricted to arise from a translationally invariant
state of $N$ spins while the optimal $\Gamma$ can automatically be
chosen so since $H_\mathrm{Wolf}$ is translationally invariant.

An important observation can be made from
Eq.~(\ref{eq:def_global_opt}), namely that as $H_\mathrm{Wolf}$ commutes
with the $z$ component of the total spin, in the considered limit of
$N\rightarrow\infty$ OW were right when they made
assumption (i): The optimal state can indeed be chosen to have a
definite number of spins ``down'' and thus does not contain 
superpositions of states with different $p$ values. Conversely, 
from our previous work
\cite{meyer} we know that Eq.~(\ref{eq:def_global_opt}) is also valid
for any fixed $p$, \ie we can write $C^\text{max}(N,p)$ on the
left-hand side when making the appropriate restrictions on $\Gamma$. In
summary, for fixed $p$ the maximal concurrence is given by:
\begin{equation}
  \label{eq:con_of_Egs}
  C^\text{max}(N,p)=-\inf_s E_\text{GS}[H_\text{Wolf}(s),p],
\end{equation}
where $E_\text{GS}[H_\text{Wolf}(s),p]$ is the ``ground state'' energy of
$H_\text{Wolf}(s)$ in the manifold of states with $p$ spins
``down''. The overall maximal concurrence is given by further
optimization over $p$ or, equivalently, by using unrestricted
ground state energies:  
\begin{equation}
  \label{eq:unres_con}
  \begin{split}
    C^\text{max}(N)
    &=
    \max_p C^\text{max}(N,p)
    \\
    =&
    -\inf_s E_\text{GS}[H_\text{Wolf}(s)].
  \end{split}
\end{equation}


Let us now describe the connection with the work of Yang and Yang.  In
their seminal papers Yang and Yang
\cite{yang66:_one_dimen_I,yang66:_one_dimen_II,yang66:_one_dimen_III}
study this anisotropic Heisenberg XXZ Hamiltonian (see
e.g.~\cite{korepin93:_quant_inver_scatt} for more recent work):
\begin{equation}
  \label{eq:H_yang}
  H_\text{Yang} 
  =
  - \frac{1}{2}\sum_i 
  \left\{
    \sigma_x^{i}\sigma_x^{i+1}+\sigma_y^{i}\sigma_y^{i+1}
    + \Delta \sigma_z^{i}\sigma_z^{i+1} 
  \right\},
\end{equation}
and they define $f=\lim_{N\rightarrow\infty}f_N$, where $f_N$ is half
the energy per spin in the ground state with a given number $p$ of
spins ``down'':
\begin{equation}
  \label{eq:def_f}
  f_N(\Delta,y)
  =
  \frac{1}{2N}E_\text{GS}(H_\text{Yang},p)
  .
\end{equation}
Here $y$ is the average magnetization:
\begin{equation}
  \label{eq:def_y}
  y=\frac{1}{N}\langle \sum \sigma_z^i \rangle 
  = 1-\frac{2p}{N}.
\end{equation}
Since $p$ is a conserved quantum number, one can include a magnetic field
along $z$ and only shift the energy of each eigenstate. 
The translation of the results of Yang and Yang to our
optimization problems is therefore,
\begin{multline}
  \label{eq:H_from_E}
  E_\text{GS}[H_\text{Wolf}(s),p]    
    =
    \\
  2 f_N(\Delta(s),y)-\mathcal{H}(s) y - \frac{1}{2} \Delta(s),
\end{multline}
with $s>0$. Note that $\Delta^2-\mathcal{H}^2=1$. 
To find $C^\text{max}(N,p)$ we
should minimize Eq.~(\ref{eq:H_from_E}) over $s$ while keeping $y$ fixed at
the value corresponding to $p$ [cf.\ Eq.~(\ref{eq:def_y})]. To find the
overall maximal concurrence $C^\text{max}(N)$ we should furthermore
minimize over $y$.

\section{The  phase-diagram of the XXZ model}
\label{sec:phasediagram}

We are considering the XXZ model in the limit
$N\rightarrow\infty$. The second Yang and Yang paper 
\cite{yang66:_one_dimen_II}
deals with the properties of $f(\Delta,y)$ in exactly this limit. The
third paper \cite{yang66:_one_dimen_III} contains information about
the magnetic properties, i.e. it is highly relevant when we also vary
$y$ in order to find the optimal fraction of spins ``down''.
\begin{figure}[tbp] 
  \centering
  \resizebox{8.5cm}{!}{
    \includegraphics{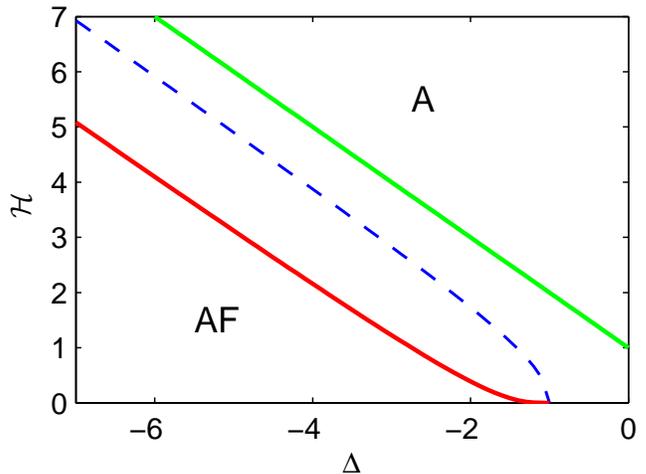}
    }
    \caption{Phase diagram for the XXZ chain in a
      magnetic field. For large negative $\Delta$ and small magnetic
      field $\mathcal{H}$ the ground state has perfect
      anti-ferromagnetic ordering (AF) with vanishing magnetization. When
      the magnetic field is increased beyond a certain critical value
      (lower bold line in the plot), a non-vanishing magnetization
      develops. The magnetization increases with increasing field
      until finally all the spins are aligned. This aligned phase (A) is
      entered at an upper critical field indicated by the upper fat
      line in the plot. We are interested in the properties of the
      chain along the dashed line in the plot, i.e. we deal with
      the phase with non-vanishing, but also non-saturated
      magnetization.}
  \label{fig:phasediagram}
\end{figure}
In order to understand the regimes of parameters we are
interested in and relate them to the known properties of the
model, it is useful to look at the phase
diagram of this model, displayed  in Fig.~\ref{fig:phasediagram}.

Let us first identify the region of the phase diagram which belongs
to our parent Hamiltonian (\ref{eq:def_H}). From
Eq.~(\ref{eq:def_Delta_and_mathcalH}) it is clear that as $s$ varies from $0$
to $\infty$, we move on a hyperbola in the $\Delta$--$\mathcal{H}$
plane: The $s=0$ case corresponds to $(-\infty,\infty)$, whereas at
$s=1$ we are at the point of closest approach and cross the $\Delta$
axis in $(-1,0)$, and as $s \rightarrow \infty$ we move back to
infinity, but this time with negative magnetic field. Comparing this with Fig.
\ref{fig:phasediagram}, it is then easy and not surprising to see that
the $s$-hyperbola lies exactly in the ``difficult'' region of the
phase diagram, i.e., the part with neither perfectly aligned spins nor
perfect anti-ferromagnetic order (between AF and A in Fig.
\ref{fig:phasediagram}).

Since a change of sign of the magnetic field will only interchange the
role of spin ``up'' and spin ``down'', we can ignore the negative
$\mathcal{H}$ branch and focus on $s\in[0,1]$. Then each
point on the curve corresponds to exactly one $\Delta$ and we can thus
parameterize the curve by $\Delta$ instead of $s$. The optimization
is then done over $\Delta$ with the magnetic field always given by
$\mathcal{H}=\sqrt{\Delta^2-1}$. 

\section{The integral equation}
\label{sec:int_eq}

The Bethe ansatz basically consists of the assumption that the wave
function can be written as a sum of plane waves with a limited number
of terms.  If we are looking for a state with $p$ spins ``down'', only
$p$ wave numbers are needed. For the XXZ chain the first Yang and Yang
article shows that this is indeed enough to produce the ground state
wave function~\cite{yang66:_one_dimen_I}. For our purposes we should
note that Yang and Yang give explicitly the equations one needs to
solve in order to find the ground state energy. In the
limit of $N\rightarrow\infty$, the number of wave numbers naturally
becomes infinite and the equation to find them becomes an integral
equation for the wave number density. This equation mathematically has
a form of a so called Fredholm equation of the second kind. After some
reparameterization this equation attains the form (Eq. [7a] in Ref.
\cite{yang66:_one_dimen_II}):
\begin{equation}
\label{eq:int_eq}
R(\alpha)
=
\frac{dp}{d\alpha}
-
\frac{1}{2\pi}
\int_{-b}^{b}
\frac{\partial \theta}{\partial \beta} R(\beta) 
d\beta
.
\end{equation}
The unknown function here is $R$, which is the reparameterized density
of wave numbers.  The other functions depend parametrically on
$\Delta$, and in terms of the parameter $\lambda=\cosh^{-1}(-\Delta)$,
they are explicitly given by:
\begin{align}
\label{eq:def_dpda}
\frac{dp}{d\alpha}
&=
\frac{\sinh\lambda}{\cosh\lambda-\cos\alpha}
,
\\
\label{eq:def_dtdb}
\frac{\partial \theta}{\partial \beta}
&=
\frac{\sinh2\lambda}{\cosh2\lambda-\cos(\alpha-\beta)}
.
\end{align}
Let us point out the importance of the integration limit $b$ in
Eq.~(\ref{eq:int_eq}): When varying $b$, we get solutions
corresponding to different values of $y$. In fact, $y$ is given by:
\begin{equation}
  \label{eq:int_y}
  \pi(1-y)=\int_{-b}^b R(\alpha) d\alpha
  .
\end{equation}
Note, however, that $R$ also depends on $b$, so the connection is
not very obvious. In praxis (i.e.\ when doing numerics) one solves
Eq.~(\ref{eq:int_eq}) for a range of the parameter $b$ in order to 
find the result for the wanted values of $y$. If one wants to optimize 
some quantity with respect to $y$, however, this can equally well  be
achieved by optimizing with respect to $b$.

We are not primarily interested in $R$ (which describes the state),
but in $f$, which is the energy. It is given by:
\begin{equation}
  \label{eq:int_f}
  f(\Delta,y)
  =
  -\frac{\Delta}{4}
  -\frac{\sinh\lambda}{2\pi}\int_{-b}^{b}R(\alpha) \frac{dp}{d\alpha} d\alpha
  .
\end{equation}
Again, $f$ is written as a function of $y$, but in praxis the
dependence is via $b$.

\section{Numerical solution of  the integral equation}
\label{sec:solving}

A possible way to solve Eq.~(\ref{eq:int_eq}) is to turn
the integral into a sum so that it becomes a matrix equation. This is
called the \emph{Nystrom method}~\footnote{See e.g. Numerical
  Recipes.}. The best way to discretize an integral is \emph{not}
always equally spaced points; very often it is much more efficient  to use a  
\emph{Gaussian Quadrature}. This means that we evaluate the integrand
at $M$ points $\{\alpha_k\}$ and make a weighted sum with weights
$\{w_k\}$. The points and the weights can be easily found  in e.g.\
Mathematica. In this way, Eq.~(\ref{eq:int_eq}) becomes:
\begin{equation}
  \label{eq:mat_eq}
  R_k=\xi_k-\sum\limits_l w_l K_{kl}R_l
\end{equation}
where
\begin{equation}
  \label{eq:def_dis}
  R_k=R(\alpha_k) 
  \quad,\quad
  \xi_k=\frac{\sinh\lambda}{\cosh\lambda-\cos\alpha_k},
\end{equation}
while
\begin{equation}
  K_{kl}=\frac{1}{2\pi}
  \frac{\sinh2\lambda}{\cosh2\lambda-\cos(\alpha_k-\alpha_l)}
  .
\end{equation}
It is clear that Eq.~(\ref{eq:mat_eq}) is a matrix equation and that
solving it cannot be harder than inverting $\Id+\tilde{K}$ where
$\tilde{K}_{kl}=w_l K_{kl}$ (no summation over $l$).

The advantage of using Gaussian Quadrature is that one does not need
too many points to get a very good estimate of the integral for any
sensible function. What exactly a ``sensible function'' is depends on
the exact Gaussian Quadrature rule used. We use the simple
Gauss-Legendre rule, assuming that $R$ is well approximated by a
polynomial on the interval $[-b,b]$. This is reasonable here because
(\ref{eq:def_dpda}) and (\ref{eq:def_dtdb}) are well-behaved for the
values of $\lambda$ we will consider. The final matrix equation can
be solved very rapidly on a small size computer. A moderate value of
$M$, however, means that our knowledge of $R$ is restricted to a
rather crude sampling; fortunately this is not a problem, since $y$
and $f$ are themselves integrals, and so can be evaluated with the
full accuracy of Gaussian Quadrature.

\begin{figure}[tbp]
  \centering
  \resizebox{8cm}{!}{
    \includegraphics{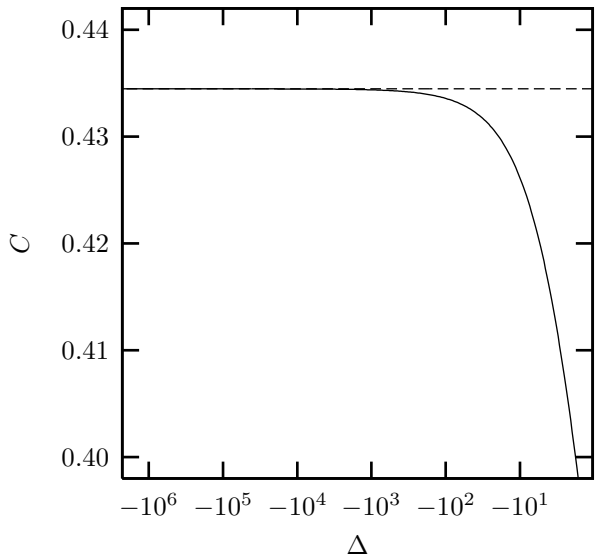}
    }
    \caption{We plot $-E_\text{GS}[H_\text{Wolf}]$, that is,
      for each $\Delta$ the optimal $y$ is chosen. It can be seen that
      the maximum is attained in the limit $\Delta\rightarrow -\infty$
      and that the limiting value coincides with
      OW's result indicated by the dashed line.}
  \label{fig:H_of_d}
\end{figure}

To give the reader an idea about the numerics, we note that a simple
Mathematica program will work very well with $M\le 30$. To produce a
plot $f(\Delta,y)$ versus $y$ for $\Delta$ not too close to $-1$
it takes about one minute. To plot the function of main interest,
Eq.~(\ref{eq:H_from_E}) optimized over $p$ (i.e.\ $y$, i.e.\ $b$) also
only takes a few minutes. In Fig.~\ref{fig:H_of_d} we present the
results of a Fortran program, which is (not surprisingly) much faster
than the initial Mathematica code. The results indicate that
OW's assumption (ii) was correct: When we plot
$E_\text{GS}[H_\text{Wolf}]$ as function of $\Delta$, we see that the
optimal value of $\Delta$ is reached at $-\infty$, and in this limit
OW's result is recovered.  We conclude that these
simple numerical results indicate that the state that maximizes the NN
concurrence without any restrictions (i.e.  optimized over $p/N$, i.e.
$y$) coincides with the OW state fulfilling assumption (ii) 
(no NN pairs of spins ``down'').

\section{Perturbative calculation}
\label{sec:pert_calc}

Looking at Eq.~(\ref{eq:H_from_E}) above we see that the finite value
in Fig.~\ref{fig:H_of_d} in the limit $\Delta \rightarrow -\infty$ is
obtained because some diverging terms happen to cancel each other.
This is of course a great concern when doing numerics since it means
that a good relative precision (knowing the result to e.g.\ 1 ppm) may
not be enough. The obvious strategy is to extract the solution in the
strict limit $\Delta\rightarrow-\infty$.  In this section, we will 
present a perturbative calculation in $1/\Delta$.

In zeroth order of the perturbation series we set $\cosh\lambda=\infty$
in Eq.~(\ref{eq:int_eq}), and arrive at the simple equation:
\begin{equation}
  \label{eq:int_eq_limit}
  R_0(\alpha) = 1 - \frac{1}{2\pi} \int_{b}^{b} R_0(\beta) \; d\beta 
  .
\end{equation}
The right hand side does not depend on $\alpha$ and we easily find the
constant solution:
\begin{equation}
  \label{eq:R0}
  R_0=\frac{1}{1+\frac{b}{\pi}}
  \quad\text{and}\quad
  y_0=\frac{1-\frac{b}{\pi}}{1+\frac{b}{\pi}}.
\end{equation}
This means that in this limit $f=-\Delta f_{-1}$ with:
\begin{equation}
  \label{eq:f0}
  f_{-1} 
  = 
   \frac{1}{4} - \frac{\frac{b}{\pi}}{1+\frac{b}{\pi}}
  =
  - \frac{1}{4} + \frac{1}{2} y_0
\end{equation}
and that Eq.~(\ref{eq:H_from_E}) thus gives 0, independently of $y_0$.
At this level of precision we therefore get no information as
to whether OW's solution is optimal for all $y$'s.

The next order is ``$1/\Delta$'', i.e.\ we expand both sides of
Eq.~(\ref{eq:int_eq}) and equate terms proportional to
$-1/\Delta=1/\cosh\lambda$. We get
\begin{equation}
  \label{eq:int_order1}
  R_1(\alpha)
  =
  \cos\alpha
  -\frac{1}{2\pi}\int_{-b}^b R_1(\beta) \; d\beta
  .
\end{equation}
The $\alpha$ dependence on the right hand side is
$\cos\alpha$ plus a constant, so we easily find
\begin{equation}
  \label{eq:R1}
  R_1(\alpha)= \cos\alpha - \frac{\sin b}{\pi+b}
  \quad\text{and}\quad
  y_1=-\frac{2\sin b}{\pi+b}
  .
\end{equation}
The correction to $f$ is given by
\begin{equation}
  \label{eq:f1}
  \begin{split}
  f_0=&-\frac{1}{2\pi}\int_{-b}^b \left[-\cos\alpha R_0 +
    R_1(\alpha)\right]\;d\alpha
  \\
  =& -\frac{ 2 \sin b}{b+\pi}
  \\
  =& y_1
  .
  \end{split}
\end{equation}
This means that in Eq.~(\ref{eq:H_from_E}) we get a zeroth order
contribution of
\begin{equation}
  \label{eq:H0}
  E_{\text{GS},0}=2y_1-y_1=y_1= -\frac{ 2 \sin b}{b+\pi}
  .
\end{equation}
It is easy to see that this expression is the same as the one obtained
by OW in Ref.~\cite{Wootters} and if we do a
numerical optimization over $b$ we arrive at the notorious
$0.434467\ldots$ for the maximal concurrence. This value is obtained
for $b=b_\text{OW}= 1.351802\ldots$, corresponding by
Eq.~(\ref{eq:int_y}) to $y=y_\text{OW}= 0.398316\ldots$.

\subsection{Recursion formula for higher order corrections}
\label{sec:recursion_formula}

It is tedious, but essentially not difficult to continue in the above
fashion and calculate higher order corrections. A useful trick is to
develop a recursion formula. Let us write $\epsilon=1/|\Delta|$ and
define:
\begin{equation}
  \label{eq:def_power_series}
  \begin{split}
    R(\alpha) &= \sum_k R_k(\alpha) \; \epsilon^k\\
    \frac{d p(\alpha)}{d\alpha} 
    & = 
    \sum_k \frac{d p_k(\alpha)}{d\alpha} \; \epsilon^k
    \\
    \frac{\partial \theta(\alpha,\beta) }{\partial\beta}
    & = 
    \sum_k \frac{\partial \theta_k(\alpha,\beta)}{\partial\beta} \; \epsilon^k
    .
  \end{split}
\end{equation}
The $k$'th order terms of Eq.(\ref{eq:int_eq}) give us:
\begin{multline}
  \label{eq:kth_order_int_eq}
  R_k(\alpha) 
  = 
  \\
  \frac{dp_k(\alpha)}{d\alpha}
  -\frac{1}{2\pi} \int_{-b}^{b}
  \sum_{j=0}^k
  \frac{\partial\theta_{k-j}(\alpha,\beta)}{\partial\beta} 
  R_{j}(\beta) \; d\beta
  .
\end{multline}
Using the fact that $\frac{\partial\theta_0}{\partial\beta}=1$ we collect terms
containing $R_k$ on the left hand side:
\begin{equation}
  \begin{split}
    \label{eq:Rk_on_lhs}
    \int_{-b}^{b} \Big[
      \delta(\alpha-&\beta)+\frac{1}{2\pi}
    \Big] R_k(\beta) \; d\beta
  \\
    = &
    \frac{p_k}{d\alpha}
    -\frac{1}{2\pi} 
    \int_{-b}^{b}
    \sum_{j=0}^{k-1} \frac{\partial\theta_{k-j}(\alpha,\beta) }{\partial\beta}
    R_{j}(\beta) 
    \; d\beta
    \\
    = &
    q_k(\alpha)
    ,
  \end{split}
\end{equation}
where we have introduced $q_k(\alpha)$ as a shorthand notation for the
r.h.s. The r.h.s.\ depends only on the known functions $dp/d\alpha$
and $\partial\theta/\partial\beta$, and on $R_{j}$ for $j<k$. The
integral operator acting on $R_k$ on the l.h.s. of
Eq.(\ref{eq:Rk_on_lhs}) can easily be inverted since it is built from
the identity and a projection operator (onto a constant). We finally
end up with the recursion formula
\begin{equation}
  \label{eq:Rk_recursion}
  R_k(\alpha)
  =
  q_k(\alpha)
  -\frac{1}{2}\frac{1}{b+\pi}\int_{-b}^{b}q_k(\beta)d\beta
  .
\end{equation}
In terms of $R_k$ and the auxiliary function $q_k$, we have for
$y_k$, $k>0$:
\begin{equation}
  \label{eq:yk}
  y_k=2\left[R_k(\alpha)-q_k(\alpha)\right]
  .
\end{equation}
Note that despite the appearence of $\alpha$ on the right-hand side,
this relation does make sense since the form of
Eq.(\ref{eq:Rk_recursion}) ensures that only terms independent of
$\alpha$ survive.

Using Eq.(\ref{eq:Rk_recursion}), it is fairly easy to show that
\begin{equation}
  \label{eq:R2}
  \begin{split}
    R_2(\alpha)
    =&
    \cos^2\alpha
    -\frac{\sin b}{b+\pi}\; \cos\alpha 
    \\
    &- \frac{1}{2}\frac{\sin b}{b+\pi}
    \left(
      \cos b -\frac{\sin b}{b+\pi}
    \right)
    -\frac{1}{2}
  \end{split}
\end{equation}
and thus
\begin{equation}
  \label{eq:y2}
  y_2=-\frac{\sin b}{b+\pi}\left(\cos b -\frac{\sin b}{b+\pi}\right)
  .
\end{equation}
Calculating the first order contribution to the ground state energy we find the expression:
\begin{equation}
  \label{eq:H1}
  E_{\text{GS},1}(b)
  =
  \frac{1}{2}
  -\frac{b}{\pi}
  -\frac{1}{\pi}\frac{\sin b}{b+\pi}
  \left[\left(b+2\pi\right)\cos b - 2\sin b\right]
  .
\end{equation}

\subsection{Derivative at fixed $y$}
\label{sec:deriv_fix_y}

\begin{figure}[tbp]
  \centering
  \resizebox{8cm}{!}{
    \includegraphics{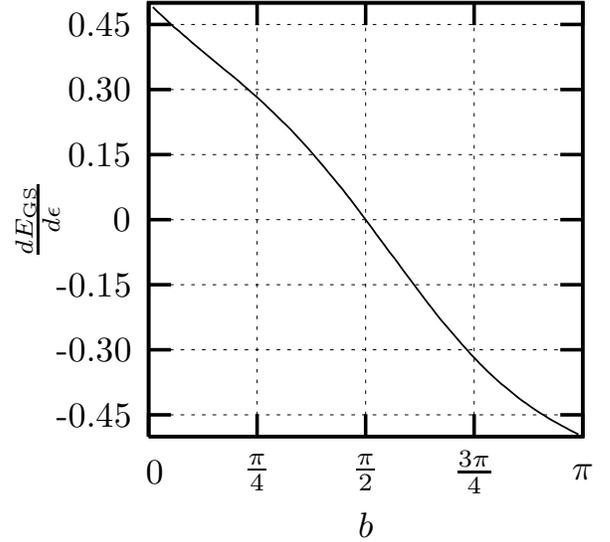}
  }
  \caption{Derivative of $E_\text{GS}$ w.r.t.\ $\epsilon=1/|\Delta|$ for
    $\epsilon=0$ and fixed $y$. A positive value for a given $b$
    indicates that OW's solution
    is a local minimum for the corresponding $y$. Negative values
    indicates that OW's solution is not a local, and thus
    also not a global minimum.}
  \label{fig:dHde_fixed_y}
\end{figure}

As mentioned above, $ E_{\text{GS},1}$ gives us access to whether
OW's solution is at least a \emph{local} minimum for a
given $y$. In Eq.(\ref{eq:H1}), $ E_{\text{GS},1}$ is expressed as a
function of $b$, so in order to calculate the derivative at fixed $y$ we
need to use the appropriate implicit differentiation rule. Calculating
the lowest non-vanishing order we find:
\begin{equation}
  \label{eq:dHde_fix_y}
  \begin{split}
    \left( \frac{dE_\text{GS}}{d\epsilon} \right)_{\!\!y}
    =&
    \frac{\partial E_\text{GS}}{\partial \epsilon}
    +
    \frac{\partial E_\text{GS}}{\partial b}
    \left(\frac{\partial b}{\partial \epsilon}\right)_{\!\! y}
    \\
    =&
    E_{\text{GS},1}(b)-\frac{d  E_{\text{GS},0}}{db}
    \frac{y_1}{d y_0/d b}+O(\epsilon)
    \\
    =&
    \frac{1}{2}-\frac{b}{\pi}+\frac{ b \; \sin b \; \cos b}{\pi (\pi +b)}
    +O(\epsilon).
  \end{split}
\end{equation}
Again we end up with a somewhat complicated expression, so we plot its
graph in Fig.~\ref{fig:dHde_fixed_y}.
We note that
$(dE_\text{GS}/d\epsilon)_y$ is positive for low $b$, but already at
$b=\pi/2$ (corresponding to $y=1/3$) it changes
sign and becomes negative. This means that for higher $b$'s, i.e.
lower $y$'s, OW's solution cannot be optimal as it is not
even a local minimum.

We conclude that in the region of sufficiently large magnetizations,
i.e. $y\ge 1/3$, the OW states (with no NN pairs of spins
``down'') maximize the NN entanglement locally, i.e. we cannot
increase the NN entanglement by allowing \emph{small} admixtures of
states with NN pairs of spins ``down''. For smaller magnetizations,
i.e.  $0\le y< 1/3$, the states that maximize the NN
entanglement necessarily contain NN pairs of spins ``down''.

\subsection{Higher orders}
\label{sec:high_order}

\begin{figure}[tbp]
  \centering
  \resizebox{8.5cm}{!}{
    \includegraphics{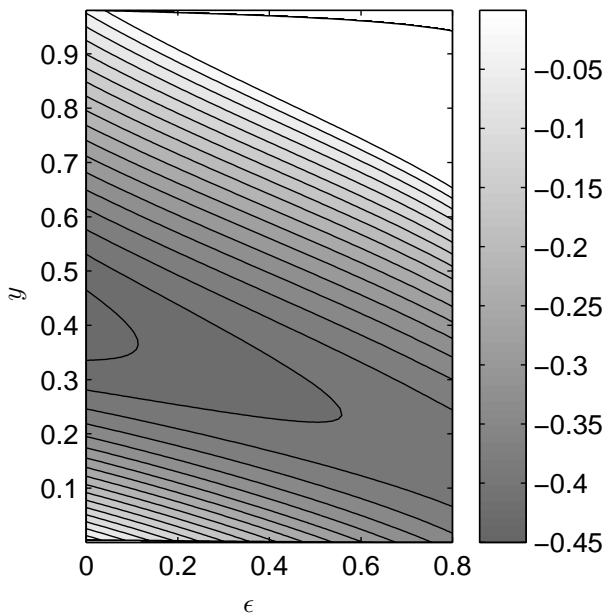}
  }
  \caption{$H_\text{Wolf}(\epsilon,y)$ calculated numerically from the
    recursion formula (\ref{eq:Rk_recursion}). For each $b$ we find
    the expansion coefficients of $H_\text{Wolf}$ and $y$ up to the fourteenth
    order in $\epsilon=1/|\Delta|$. The overall optimal state has
    $y=0.398316\ldots$. From Sec.~\ref{sec:deriv_fix_y} we know that
    for all $y>1/3$, $\epsilon=0$ is a local minimum for
    $H_\text{Wolf}$ and this plot indicates that it is also a global one.}
  \label{fig:H_num}
\end{figure}

The recursion formula (\ref{eq:Rk_recursion}) is also well suited for
numerical calculations. In Fig.~\ref{fig:H_num} we show a contour-plot
based on such a calculation including all terms up to fourteenth order in
$\epsilon=1/|\Delta|$. The plot indicates that the calculation in
Sec.~\ref{sec:deriv_fix_y} gives the global answer, i.e., for
all $y \ge 1/3$ the optimal state has no neighboring spins ``down''.

Since we perform here the perturbative calculation up to the 14th
order, we expect that this calculation allows us also to obtain some
information about the region of $y < 1/3$. From the Fig.
\ref{fig:H_num} (or more precisely from the numerical data), one can
read off the optimal value of $\epsilon$, i.e. optimal value of
$\Delta$. Solving the Bethe ansatz integral equation for this value of
$\Delta$ we can recover in this way the full information about the
corresponding optimal quantum state.

\section{Conclusions}
\label{sec:conclusions}
 
In this paper we have studied the question posed by O'Connor and
Wootters concerning translationally invariant states of $N$ qubits
with maximal nearest neighbor (NN) concurrence. We have answered this
question for $N\rightarrow\infty$ using the mapping of the problem
onto the search for ground states of a certain family of ``parent''
Hamiltonians, described by the XXZ model.  Using the analytic Bethe 
ansatz solutions of the XXZ model in the limit $N\to \infty$ (combining
analytic results of low order perturbation theory and a numerical
calculation of the 14th order perturbation theory) we have proved
that: (i) for a given number of spins ``down'', i.e. a given
magnetization $y$ larger than $1/3$, the states that
maximize the NN concurrence coincide with the ones
obtained by O'Connor and Wootters, i.e. do not have NN pairs of spins
``down''; (ii) For small magnetizations, more explicitly for $0\le
y\le 1/3$, the states that maximize the NN concurrence do
contain nearest neighbor pairs of spins ``down''; (iii) in particular,
the state that maximizes the NN concurrence without constraint on $y$
belongs to the family introduced by O'Connor and Wootters. Our results
shed more light on the subtle relations between entanglement in spin
1/2 models and the ferromagnetic/anti-ferromagnetic character of
spin-spin interactions.  In the appendix we present some simple bounds
on the optimal magnetic field that corresponds to the maximal NN
concurrence.

We acknowledge the support of The Danish Natural Science Research
Council, DFG (SFB 407, SPP 1078, SPP 1116), ESF Program ``QUDEDIS'',
EU FET IST 6th Framework Integrated Project ``SCALA'', and Spanish MEC
Grant FIS2005-04627.

\begin{appendix}

\section{Bound on the optimal magnetic field}
\label{sec:analytic}

If one keeps $y$ fixed, and use $\mathcal{H}$ to parameterize the $s$-curve
instead of $s$, one gets:
\begin{eqnarray}
  \label{eq:dhdh}
  \frac{dE_\mathrm{GS}}{d\mathcal{H}}
  &=&
  -\frac{d\Delta}{d\mathcal{H}}
  \frac{1}{2}\left[
    1+\langle \sigma_z\sigma_z\rangle
  \right]
  - y\nonumber\\
  &=& 2\frac{\mathcal{H}}{\sqrt{\mathcal{H}^2+1}} P
  +\left(\frac{\mathcal{H}}{\sqrt{\mathcal{H}^2+1}}-1\right)y,\nonumber
\end{eqnarray}
where $P$ is the probability of two
neighboring spins being both ``down''.

Demanding that  $\frac{dE_\mathrm{GS}}{d\mathcal{H}}=0$, we find:
\[
\mathcal{H}_\mathrm{opt} = \pm y/(2 \sqrt{P}\sqrt{P+y})
\]\\
where $\mathcal{H}_\mathrm{opt}$ is the optimal magnetic field. Obviously, 
 $\mathcal{H}_{opt} = 0$ iff $y = 0$. 

Using a simple bound on  $P$,
\[
P \leq (y+1)/2 \quad \text{in the limit} \quad N \rightarrow \infty,
\]
we obtain  a lower bound on $\mathcal{H}_\mathrm{opt}$:
\[
\mathcal{H}_\mathrm{opt} \geq y/(\sqrt{y+1}\sqrt{3y+1}).
\]
This bound does not work well for $y>1/3$, because it gives a finite
bound, maximized for $y=1$ when we find $\mathcal{H}_\mathrm{opt} \geq
1/(2 \sqrt{2})$, whereas we know that in this regime of $y$'s
$\mathcal{H}_\mathrm{opt} = \infty$.  For smaller values of $y < 1/3$,
both the optimal $\epsilon$ (i.e. $\Delta$, see Fig.~\ref{fig:H_num}),
as well as the optimal $\mathcal{H}_\mathrm{opt}$ attain finite
values, so that the bound might become more useful. In particular, the
results of Fig.  \ref{fig:H_num} suggest that as $y$ approaches zero,
the optimal $\epsilon$ approaches 1 more or less linearly, as
$1-(1/3)^{-1} y$, which in turn implies that the optimal $\Delta$
approaches $-1$ as $-1-3y $.  Thus for small $y$ and small $\Delta+1$,
the bound becomes $\mathcal{H}_\mathrm{opt} \ge
-\frac{1}{3}(1+\Delta)$, which already is not obvious (compare
Fig.~\ref{fig:phasediagram}).
\end{appendix}


\end{document}